\documentstyle[11pt,paspconf,epsf]{article}

\begin{document}

\title{Spatial and Velocity Biases}

\author{Anatoly Klypin, Andrey Kravtsov}
\affil{Department of Astronomy, New Mexico State University, Las
Cruces, NM 88001}

\author{Pedro Col\'in}
\affil{Instituto de Astronom\'ia, Universidad Nacional Aut\'onoma
de M\'exico, C.P. 04510, M\'exico, D.F., M\'exico}

\begin{abstract}
We give a summary of our recent studies of spatial and velocity biases
of galaxy-size halos in cosmological models. Recent progress in
numerical techniques made it possible to simulate halos in large
volumes with a such accuracy that halos survive in dense
environments of groups and clusters of galaxies. While the very central
parts of clusters still may have been affected by numerical effects,
halos in simulations look like real galaxies, and, thus, can
be used to study the biases -- differences between galaxies and the
dark matter. For the standard LCDM model we find that the correlation
function and the power spectrum of galaxy-size halos at $z=0$ are
antibiased on scales $r<5h^{-1}$Mpc and
$k\approx(0.15-30)h$Mpc$^{-1}$. The biases depend on scale, redshift,
and circular velocities of selected halos. Two processes seem to
define the evolution of the spatial bias: (1) statistical bias (or peak
bias) and merger bias (merging of galaxies, which happens
preferentially in groups, reduces the number of galaxies, but does not
affect the clustering of the dark matter). There are two kinds of
velocity bias.  The pair-wise velocity bias is
$b_{12}=0.6-0.8$ at $r< 5h^{-1}$Mpc, $z=0$. This bias  mostly
reflects the spatial bias and provides almost no information on the
relative velocities of the galaxies and the dark matter. One-point
velocity bias is a better measure of the velocities. Inside clusters
the galaxies should move slightly faster ($b_v =1.1-1.3$) than 
 the dark matter. Qualitatively this result can be understood using 
the Jeans equations of the stellar dynamics.

\end{abstract}

\section{Introduction}

The  distribution of galaxies is likely biased with respect to the dark 
matter. Therefore, the galaxies can be used to probe the matter
distribution only if we understand the bias. Although the problem of
bias has been studied extensively in the past (e.g., Kaiser
1984; Davis et al., 1985; Dekel \& Silk 1986), new data on high
redshift clustering and the anticipation of coming measurements have
recently generated substantial theoretical progress in the field. The
breakthrough in analytical treatment of the bias was the paper by Mo \& 
White (1996), who showed how bias can be predicted in the framework of
the extended Press-Schechter approximation. More elaborate analytical
treatment has been developed by Catelan et al. (1998ab), Porciani et
al.(1998), and Sheth \& Lemson (1998). Effects of nonlinearity and
stochasticity were considered in Dekel \& Lahav (1998). 

Valuable results are produced by ``hybrid'' numerical methods in which
low-resolution N-body simulations (typical resolution $\sim 20$kpc) are
combined with semi-analytical models of galaxy formation (e.g.,
Diaferio et al., 1998; Benson et al. 1999). Typically, results of these 
studies are very close to those obtained with brute-force approach of
high-resolution ($\sim 2$kpc) N-body simulations (e.g., Col\'in et
al. 1999a). This agreement is quite remarkable because the methods are
very different. It may indicate that the biases of galaxy-size objects
are controlled by the random nature of clustering and merging of
galaxies and by dynamical effects, which cause the merging, because
those are the only common effects in those two approaches.

Direct N-body simulations can be used for studies of the biases
only if they have very high mass and force resolution. Because of
numerous numerical effects, halos in low-resolution simulations do not
survive in dense environments of clusters and groups (e.g., Moore, Katz 
\& Lake 1996; Tormen, Diaferio \& Syer, 1998; Klypin et al., 1999).
Estimates of the needed resolution are given in Klypin et
al. (1999). Indeed, recent simulations, which have sufficient resolution 
have found hundreds of  galaxy-size halos moving inside
clusters (Ghigna et al., 1998; Col\'in et
al., 1999a; Moore et al., 1999; Okamoto \& Habe, 1999).

It is very difficult to make accurate and trustfull predictions of
luminosities for galaxies, which should be hosted by dark matter
halos. Instead of luminosities or virial masses we suggest to use
circular velocities $V_c$ for both numerical and observational
data. For a real galaxy its luminosity tightly correlate with the the
circular velocity. So, one has a good idea what is the circular
velocity of the galaxy. Nevertheless, direct measurements of circular
velocities of a large complete sample of galaxies are extremely
important because it will provide a direct way of comparing theory and
observations.

In this paper we give a summary of results presented in Col\'in et
al. (1999ab) and  Kravtsov \& Klypin (1999).

\section{Simulations}

We have simulated different cosmological models (Col\'in et al., 1999a),
but our main results are based on currently favored model with the
following parameters (s$\Lambda$CDM model):
$\Omega_0=1-\Omega_{\Lambda}=0.3$, $h=0.7$, $\Omega_b=0.032$,
$\sigma_8=1$. Using the Adaptive Refinement Tree code (ART; Kravtsov,
Klypin \& Khokhlov, 1997) we run a simulation with $256^3$ particles in
a 60$h^{-1}$Mpc box. Formal mass and force resolutions are
$m_1=1.1\times 10^9h^{-1}M_{\odot}$ and $2h^{-1}$kpc. Bound Density
Maximum halo finder was used to identify halos with at least 30 bound
particles. For each halo we find maximum circular velocity
$V_c=\sqrt{GM(<r)/r}$.

\section{Spatial bias}

\begin{figure}
\plottwo{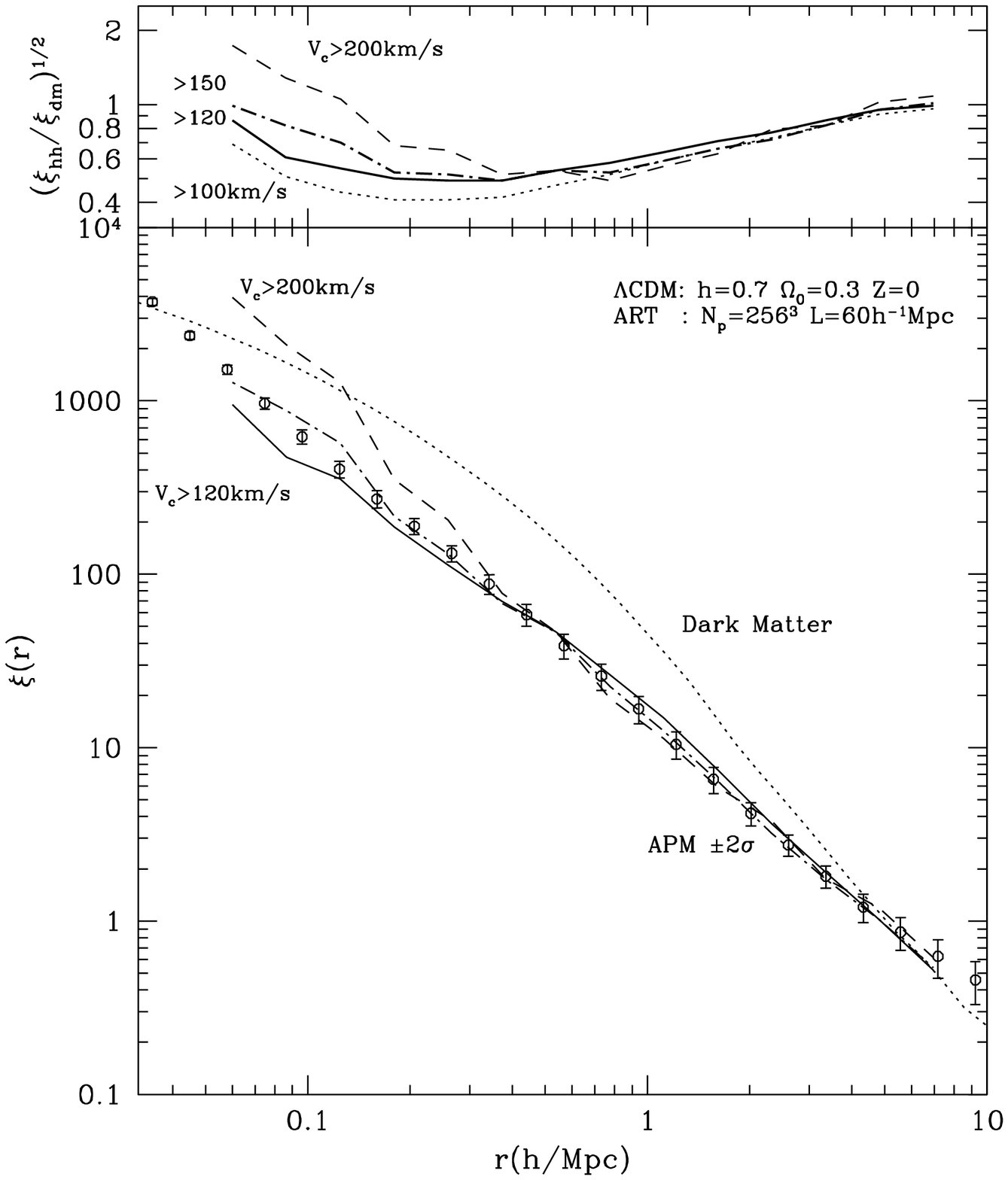}{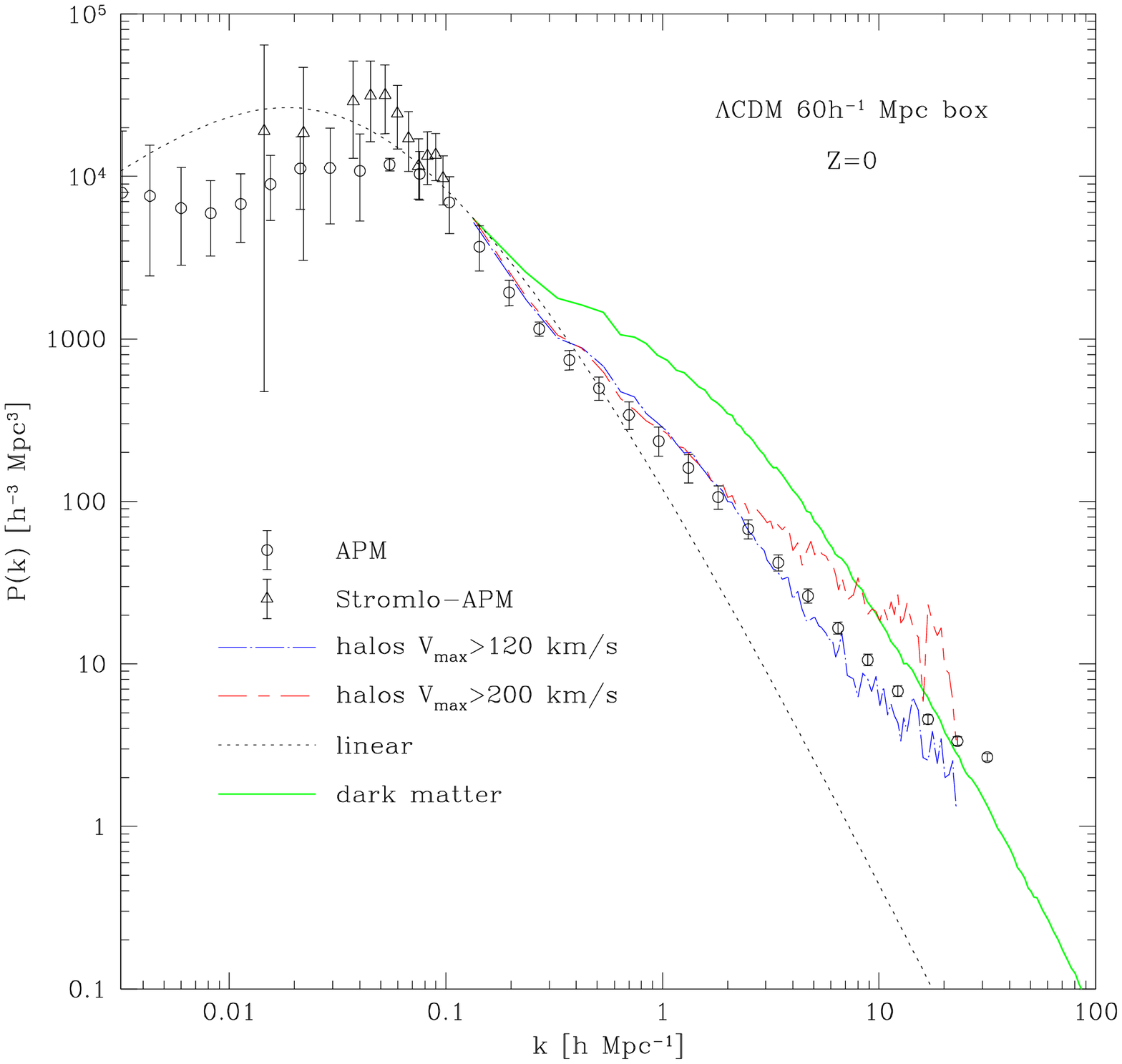}
\caption{Correlation function (left) and the power spectrum (right) of
halos of different limiting circular velocities in the $\Lambda$CDM
model. Results are compared with the observational data from the APM
and Stromlo-APM surveys.} \label{fig-1}
\end{figure}

Figure~\ref{fig-1} presents a comparison of the theoretical and
observational data on correlation functions and power spectra. The dark
matter clearly predicts much too high amplitude of clustering. The
halos are much closer to the observational points and predict
antibias. For the correlation function the antibias appears on scales $r
<5h^{-1}$Mpc; for the power spectrum the scales are $k>0.2h{\rm
Mpc}^{-1}$. One may get an impression that the antibias starts at
longer waves in the power spectrum $\lambda =2\pi/k\approx 30h^{-1}{\rm
Mpc}$ as compared with $r\approx 5h^{-1}$Mpc in the correlation
function. There is no contradiction: sharp bias at small distances in the
correlation function when Fourier transformed to the power spectrum
produces antibias at very small wavenumbers. Thus, the bias should be
taken into account at long waves when dealing with the power spectra.
There is an inflection point in the power spectrum where the nonlinear
power spectrum start to go upward (if one moves from low to high
$k$) as compared with the prediction of the linear theory. Exact
position of this point may have been affected by the finite size of the 
simulation box $k_{\rm min}=0.105h^{-1}$Mpc, but effect is expected to
be small. 

\begin{figure}
\plotone{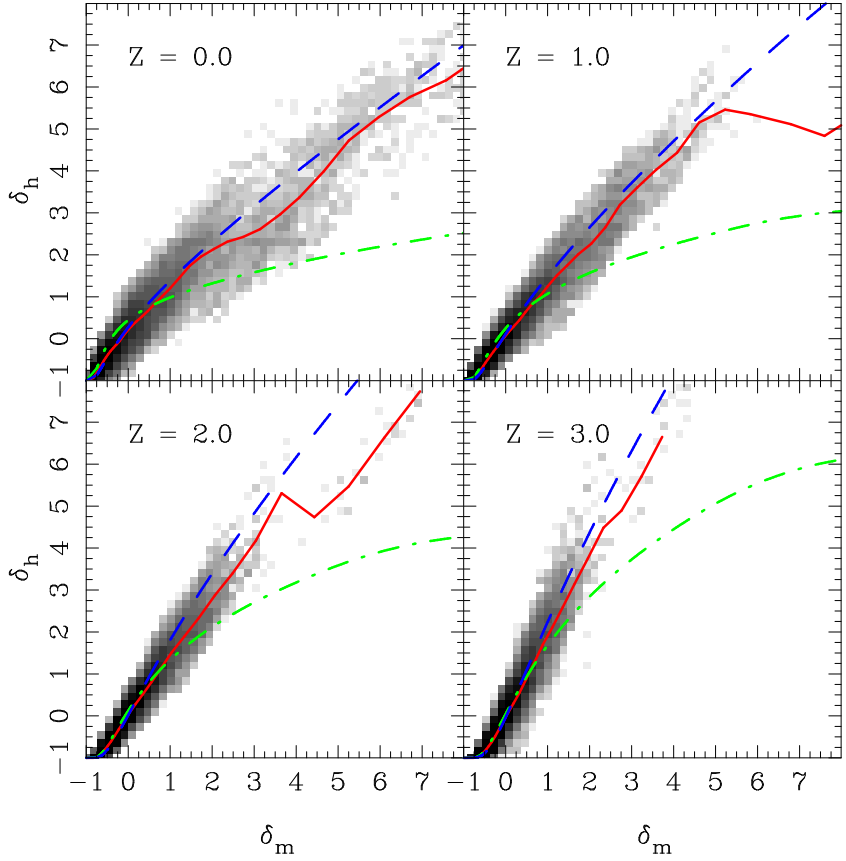}
\caption{Overdensity of halos $\delta_h$ vs. the overdensity of the
dark matter $\delta_m$. The overdensities are estimated in spheres of
radius $R_{\rm TopHat}=5h^{-1}$Mpc. Intensity of the grey shade
corresponds to the natural logarithm of the number of spheres in a 2D
grid in $\delta_h$- $\delta_m$ space. The solid curves show the average
relation. The dot-dashed curve is a prediction of an analytical model,
which assumes that formation redshift $z_f$ of halos coincides with
observation redshift (typical assumption for the Press-Schechter
approximation). The long-dashed curve is for model, which assumes that
substructure survives for some time after it falls into a larger object: 
$z_f=z+1$} \label{fig-2}
\end{figure}

At $z=0$ the bias almost does not depend on the mass limit of the
halos. There is a tendency of more massive halos to be more clustered
at very small distances $r<200h^{-1}$kpc, but at this stage it is not
clear that this is not due to residual numerical effects around centers
of clusters. The situation is different at high redshift. At very high
redshifts $z>3$ galaxy-size halos are very strongly (positively)
biased. For example, at $z=5$ the correlation function of halos with
$v_c>150{\rm km}/{\rm s}$ was 15 times larger than that of the dark
matter at $r=0.5h^{-1}$Mpc (see Fig.8 in Col\'in et al., 1999a). The
bias was also very strongly mass-dependent with more massive halos
being more clustered. At smaller redshifts the bias was quickly
declining. Around $z=1-2$ (exact value depends on halo circular
velocity) the bias crossed unity and became less than unity (antibias)
at later redshifts. 

Evolution of bias is illustrated by Figure~\ref{fig-2}. The figure
shows that at all epochs the overdensity of halos tightly correlates with
the overdensity of the dark matter. The slope of the relation depends on
the dark matter density and evolves with time. At $z>1$ halos are
biased ($\delta_h > \delta_m$) in overdense regions with $\delta_m>1$ and
antibiased in underdense regions with $\delta_m< -0.5$ At low redshifts
there is an antibias at large overdensities and almost no bias at low
densities.

\section{Velocity bias}
\begin{figure}
\plottwo{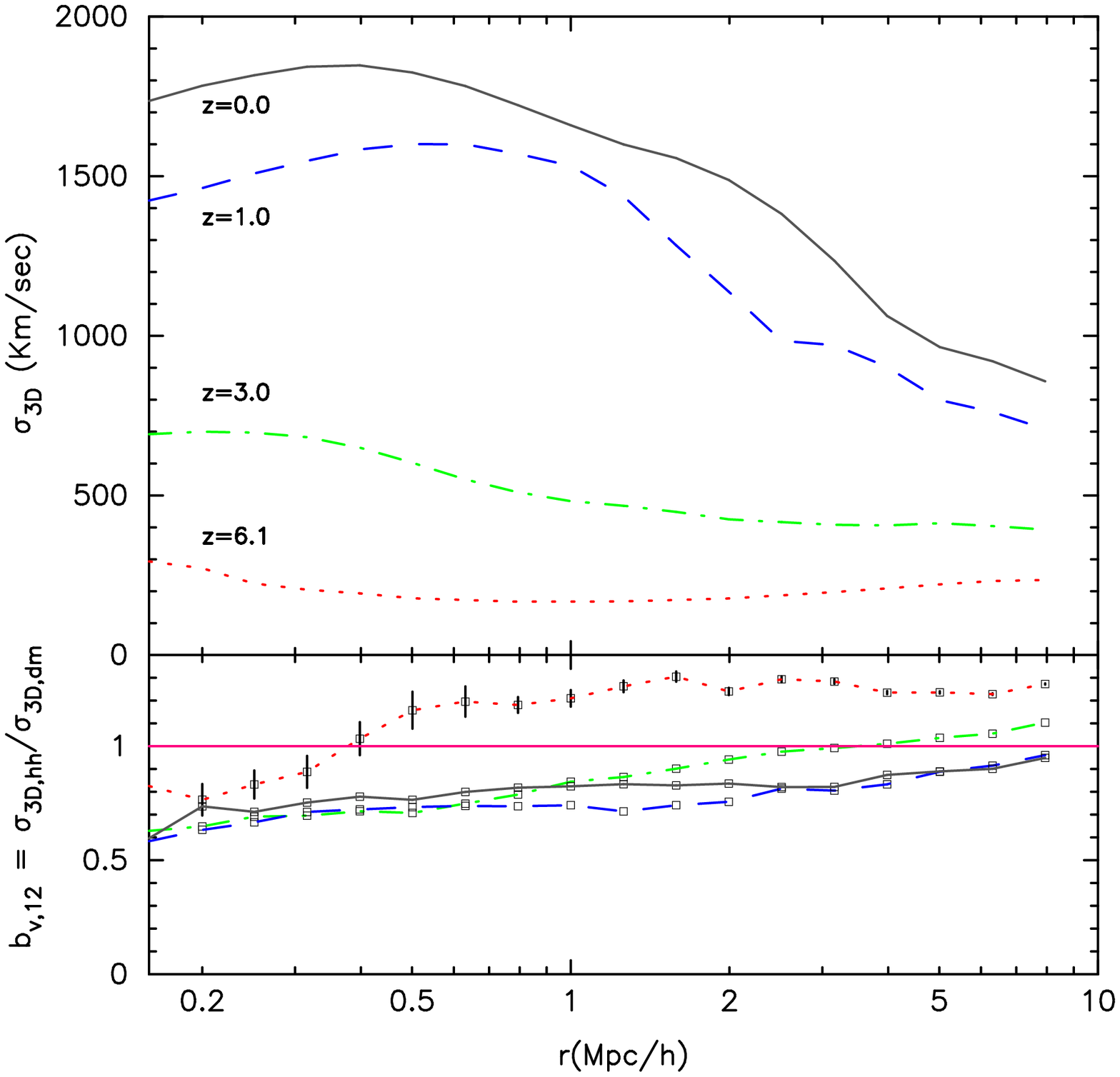}{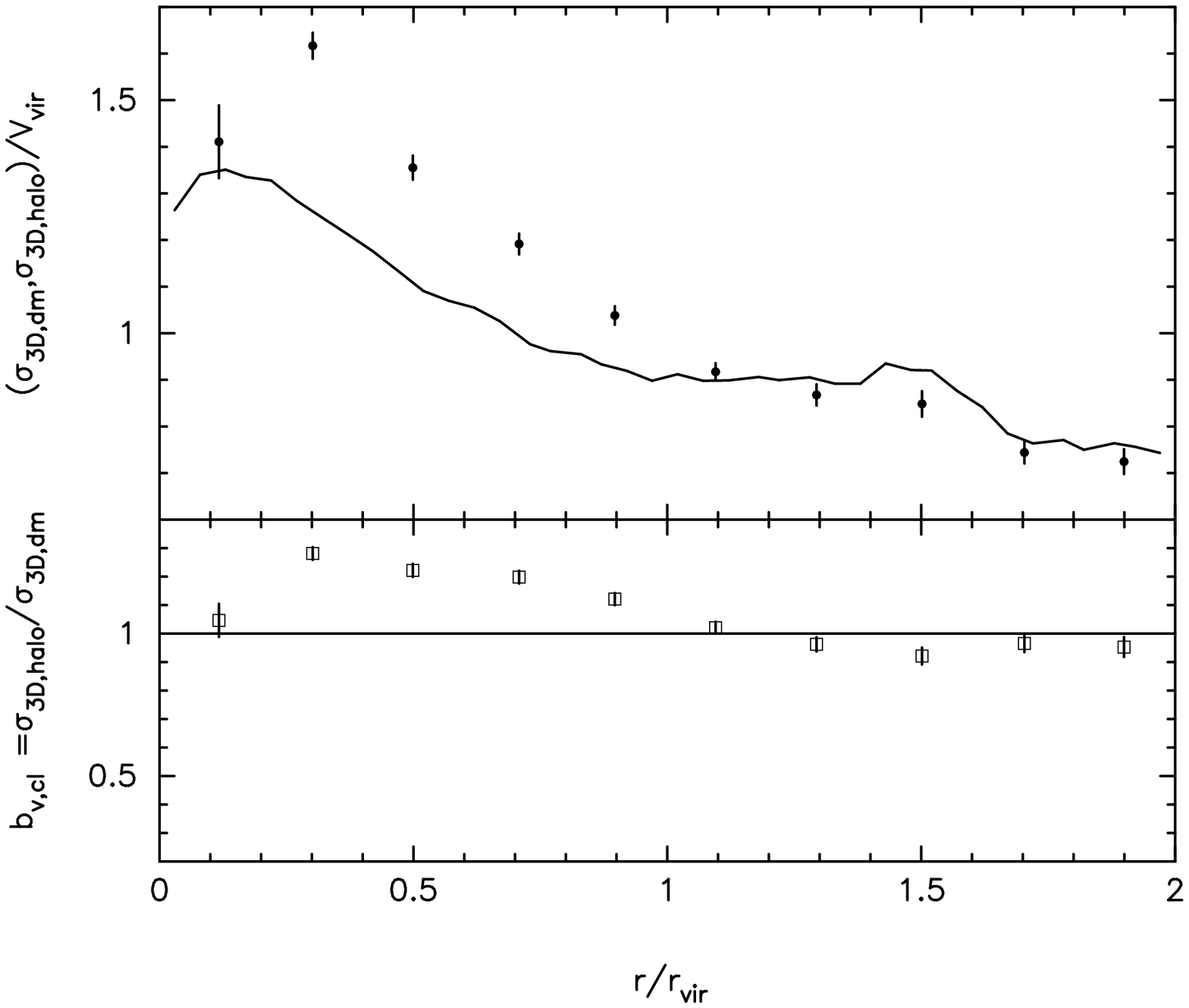}
\caption{{\it Left:} Evolution with redshift of the pairwise rms
 velocity of the dark matter (top panel) and the pairwise velocity bias
 of halos with $V_c>90$km/s (bottom panel. {\it Right:} Top panel: 3D
 rms velocity for halos (circles) and for dark matter (full curve) in
 twelve largest clusters. Bottom panel: velocity bias in the
 clusters. The bias in the first point increases to 1.2 if the central
 cD halos are excluded from analysis. Errors correspond to 1-sigma
 errors of the mean obtained by averaging over 12 clusters at two
moments of time. Fluctuations for individual clusters are larger.}
\label{fig-3}
\end{figure}

There are two statistics, which measure velocity biases -- differences
in velocities of the galaxies (halos) and the dark matter. For a review
of results and references see Col\'in et al. (1999). Two-particle or
pairwise velocity bias (PVB) measures the relative velocity dispersion in
pairs of objects with given separation $r$: $b_{12}=\sigma_{\rm
halo-halo}(r)/\sigma_{\rm dm-dm}(r)$. It is very sensitive to the
number of pairs inside clusters of galaxies, where relative velocities
are largest. Removal of few pairs can substantially change the value of
the bias. This ``removal'' happens when halos merge or are destroyed by
central cluster halos. One-point velocity bias is estimated as a ratio
of the rms velocity of halos to that of the dark matter:
$b_1=\sigma_{halos}/\sigma_{dm}$. It is typically applied to clusters
of galaxies where it is measured at different distances from the
cluster center. For our analysis of the velocity bias in clusters we
have selected twelve most massive clusters in the simulation. The most
massive cluster had virial mass $6.5\times 10^{14}h^{-1}M_{\odot}$
comparable to that of the Coma cluster. The cluster had 246 halos with
circular velocities larger than 90~km/s. There were three Virgo-type
clusters with virial masses in the range $(1.6-2.4)\times
10^{14}h^{-1}M_{\odot}$ and with approximately 100 halos in each
cluster. One cluster was excluded from the analysis  because it was in
the process of the major merger at $z=0$. 
Figure~\ref{fig-3} shows results for both statistics. Left panels show
evolution of the PVB. Just as the spatial bias, PVB is positive at
large redshifts (except for the very small scales) and decreases with
the redshift. At lower redshifts it does not evolve much and stays
below unity (antibias) at scales below $5h^{-1}$Mpc on the level
$b_{12}\approx (0.6-0.8)$.

\begin{figure}
\plotfiddle{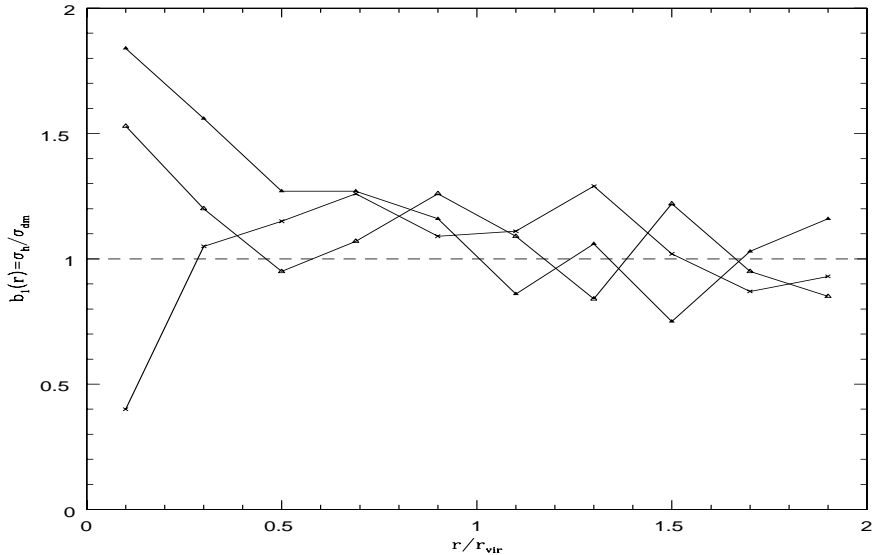}{2.2in}{0}{60}{40}{-200}{-75}
\caption{One-point velocity bias for 3 Virgo-type clusters in the
simulation. Central cD halos are not included. Fluctuations in the bias
are very large because each cluster has only $\sim$100 halos with $V_c>
90km/s$ and because of substantial substructure in the clusters. }
\label{fig-4}
\end{figure}

Right panels show one-point velocity bias
in clusters at $z=0$. Note that the sign of the bias is now different:
halos move slightly faster than the dark matter. The bias is stronger
in the central parts $b_1=1.2-1.3$ and goes to almost no bias
$b_1\approx 1$ at the virial radius and above. Both the antibias in the
pairwise velocities and positive one-point bias are produced by the
same physical process -- merging and destruction of halos in central
parts of groups and clusters. The difference is in the different
weighting of halos in those two statistics. Smaller number of
high-velocity pairs significuntly changes PVB, but it does not affect much 
the one-point bias because it is normalized to the number of halos at
a given distance from the cluster center. At the same time, merging
preferentially happens for halos, which move with smaller velocity at a 
given distance from the cluster center. Slower halos have shorter
dynamical times and have smaller apocenters. Thus, they have better
chance to be destroyed and merged with the central cD halo. Because
low-velocity halos are eaten up by the central cD, velocity dispertion
of those, which survive, is larger. Another way of addressing the issue 
of the velocity bias is to use the Jeans equations. If we have a tracer 
population, which is in equlibrium in a potential produced by mass
$M(<r)$, then
\begin{equation}
-r\sigma_r^2(r)\left[\frac{d\ln\sigma_r^2(r)}{d\ln r}+
\frac{d\ln\rho(r)}{d\ln r}+2\beta(r)\right] =GM(<r),
\end{equation}
\noindent where $\rho$ is the number density of the tracer, $\beta$ is
the velocity anisotropy, and $\sigma_r$ is the rms radial velocity. The
r.h.s. of the equation is the same for the dark matter and the
halos. If the term in the brackets would be the same, there would be no
velocity bias. But there is systematical difference between the halos
and the dark matter: the slope of the distribution halos in a cluster
$\frac{d\ln\rho(r)}{d\ln r}$ is smaller than that of the dark matter (see
Col\'in et al., 1998, Ghigna et al., 1999). The reason for the
difference of the slopes is the same --
merging with the central cD. Other terms in the equation also have
small differences, but the main contribution comes from the slope of
the density. Thus, as long as we have spatial antibias of the halos,
there should be a small positive one-point velocity bias in clusters
and a very strong antibias in pairwise velocity. Exact values of the
biases are still under debate, but one thing seems to be certain: one
bias does not go without the other.

The velocity bias in clusters is difficult to measure because it is
small. The right panel in Figure~\ref{fig-3} may be misleading because
it shows the average trend, but it does not give the level of
fluctuations for a single cluster. Note that the errors in the plots
correspond to the error of the mean obtained by averaging over 12
clusters and two close moments of time. Fluctuations for a single
cluster are much larger. Figure~\ref{fig-3} shows results for three
Virgo-type clusters in the simulation. The noice is very large because
of both poor statistics (small number of halos) and the noise produced
by residual non-equilibrium effects (substructure). Comparable (but
slightly smaller) value of $b_v$ was recently found in simulations by
Ghigna et al. (1999, astro-ph/9910166) for a cluster in the same mass
range as in Figure~\ref{fig-3}. Unfortunately, it is difficult to make
detailed comparison with their results because Ghigna et al. (1999) use
only one hand-picked cluster for a different cosmological model. Very
likely their results are dominated by the noise due to residual
substructure. Results of another high-resolution simulation by Okamoto
\& Habe (1999) are consistant with our results.

\section{Conclusions}

There is a number of physical processes, which can contribute to the
biases. In our papers we explore dynamical effects in the dark matter
itself, which result in differences of the spatial and velocity
distribution of the halos and the dark matter. Other effects related to 
the formation of luminous parts of galaxies also can produce or change
biases. At this stage it is not clear how strong are those
biases. Because there is a tight correlation between the luminosity and
circular velocity of galaxies, any additional biases are limited by the 
fact that galaxies ``know'' how much dark matter they have. 

Biases in the halos are reasonably well understood and can be
approximated on a few Megaparsec scales by analytical models. 
We find that the biases in the distribution of the halos are sufficient 
to explain within the framework of standard cosmological models the
clustering properties of galaxies on a vast ranges of scales from
100~kpc to dozens Megaparsecs. Thus, there is neither need nor much
room for additional biases in the standard cosmological model. 

In any case, biases in the halos should be treated as benchmarks for
more complicated models, which include non-gravitational physics. If a
model can not reproduce biases of halos or it does not have enough
halos, it should be rejected, because it fails to have correct dynamics 
of the main component of the Universe -- the dark matter.

\end{document}